\documentclass[twocolumn,showpacs,showkeys,preprintnumbers,amsmath,amssymb]{revtex4}

\usepackage{graphicx}
\usepackage{dcolumn}
\usepackage{bm}

\begin{document}
\title{A new mass value for $^{7}$Li}
\author{Sz. Nagy}
\email[Correspondence to: ]{Szilard.Nagy@physto.se}
\author{T. Fritioff}
\author{M. Suhonen}
\author{R. Schuch}\homepage{http://www.atom.physto.se}
\affiliation{AlbaNova University Center, Atomic Physics, S-106\hspace{0.05cm}91 Stockholm, Sweden}
\author{K. Blaum}
\affiliation{
Gesellschaft f\"ur Schwerionenforschung (GSI)
Darmstadt, Planckstra{\ss}e 1, D-64291 Darmstadt, Germany}
\affiliation{Johannes Gutenberg-Universit\"at Mainz, Staudingerweg 7, D-55128 Mainz, Germany}
\author{M. Bj\"orkhage}
\author{I. Bergstr\"om}\homepage{http://www.msi.se}
\affiliation{Manne Siegbahn Laboratory (MSL),
Frescativägen 24, S-104\hspace{0.05cm}05 Stockholm, Sweden}%

\date{\today}
\begin{abstract}
A high-accuracy mass measurement of $^{7}$Li was performed with the
\textsc{Smiletrap} Penning trap mass spectrometer via a cyclotron
frequency comparison of $^{7}$Li$^{3+}$ and H$_2^+$. A new atomic mass value of
$^{7}$Li has been determined to be  $7.016\,003\,425\,6\,(45)$\,u with a relative uncertainty of 0.63\,ppb. 
It has uncovered a discrepancy as large as 14 $\sigma$ (1.1\,$\mu$u) deviation relative to the literature value given in the Atomic-Mass Evaluation AME\,2003. 
The importance of the improved and revised $^7$Li mass
value, for calibration purposes in nuclear-charge radii and atomic
mass measurements of the neutron halos $^{9}$Li and $^{11}$Li, is
discussed.
\end{abstract}
\maketitle
The mass of an atom and its inherent connection with the atomic and
nuclear binding energy is a fundamental property of the atomic
nucleus. Accurate mass values are therefore of importance for a
variety of applications in nuclear and atomic physics studies
ranging from the verification of nuclear models and tests of the
Standard Model to the determination of fundamental constants \cite{lunn03}.
In nuclear structure studies the nuclear binding energy is the key
information and is defined as the missing mass of the bound system
$m(N,Z)$ compared to the sum of the masses of the constituent
protons $Zm_p$ and neutrons $Nm_n$:
\begin{equation}
B(N,Z) = ({Nm_n + Zm_p - m(N,Z)})c^2 \ .
\end{equation}
A most intriguing discovery in the last twenty years related to
atomic nuclei is the large nuclear matter distribution of the short lived nuclide $^{11}$Li ($T_{1/2}=8.94\,ms$) 
\cite{tan85}, which is attributed to a ``halo'' of neutrons around a
compact core of nucleons \cite{han95,JEN04,BJORN04}. A halo state
can be formed when bound states close to the continuum exist. Since
1985 a large number of high-accuracy experiments have been performed
on $^{11}$Li in order to observe the halo character also in other
nuclear ground state properties, for example in the nuclear charge
radii \cite{noer05} and in the quadrupole moment \cite{arn92} by laser spectroscopy, and in the binding energy, i.e., the neutron-separation energy via direct mass
measurements \cite{bach05}. Common to all of these experiments is the need of a
proper reference in order to calibrate the measurement device and to look for systematic uncertainties. Two of the experimental approaches, nuclear-charge radii determination and atomic mass measurements are discussed in more detail here.
Although $^{11}$Li is the best studied halo-nucleus there are only 
relatively poor and conflicting results regarding its two-neutron separation energy $S_{2n}$ \cite{bach05}. This can be resolved with an on-line Penning trap mass measurement on $^{11}$Li where the mass of $^{7}$Li reported in this article would be used for calibration purposes of the magnetic field. All on-line Penning trap mass spectrometers for short-lived radionuclides use buffer-gas filled traps or gas cells to decelerate and stop the high-energetic incoming ion beam. Thus, $^4$He$^+$ or $^{22}$Ne$^{2+}$ can not be used as calibration masses due to tremendous charge exchange losses while stopping a helium or neon beam in a helium (or neon) environment.

In general, a backbone of very well-known nuclides have been identified by the Atomic-Mass Evaluation (AME) \cite{AME03}, and high-accuracy mass values of suitable
stable nuclides are of utmost importance as mass references for
on-line mass measurements of radionuclides such as those performed at
different radioactive beam facilities worldwide \cite{bollen01}.

A high mass accuracy is also required for a determination of the nuclear-charge 
radii of the lithium isotopes $^{6,7,9,11}$Li via a
measurement of the optical isotope shift employing laser
spectroscopy \cite{noer05,Bush03,Ewald04}. The isotope shift receives
contributions from two sources: The mass shift due to the change of
nuclear mass and the field shift due to the change of nuclear-charge 
radii. Since the mass shift is much larger than the field
shift, and in order to extract the difference of charge radii,
relating often back to the stable isotopes, one has to know the
atomic structure and the nuclear masses with high accuracy.

The literature mass value of $^7$Li has a relative uncertainty of 11\,ppb \cite{AME03}. It has been derived from two input data, the mass of $^6$Li measured with an uncertainty of 2.7\,ppb in a Penning trap \cite{Dunn01} and the $Q$-value of the $^6$Li(n, $\gamma$)$^7$Li reaction with 80\,eV uncertainty \cite{AME03}. However a different $Q$-value has been reported in the literature with 90\,eV	uncertainty \cite{kok:85}, which would result in a greater than 100\,ppb different $^7$Li mass.
 
With the Penning trap mass spectrometer \textsc{Smiletrap}
\cite{Ber02} the mass of $^{7}$Li has been measured with a relative
uncertainty of 0.63\,ppb by comparing the cyclotron
frequencies of $^{7}$Li$^{3+}$ and H$_2^+$. A large deviation of 14$\sigma$ from
the literature mass \cite{AME03} has been observed,
having a not negligible effect, \emph{e.g.}, on the determination of the
nuclear charge radius. In order to find the reasons for the deviation
and to look for systematic effects the mass of $^4$He$^{2+}$ and $^6$Li$^{3+}$ have also been measured.

\textsc{Smiletrap} is a double Penning trap mass spectrometer
located at the Manne Siegbahn Laboratory in Stockholm. 
Our facility has been described in detail elsewhere
\cite{Ber02}, thus only a brief description shall be given here relevant for
the measurement of the $^{7}$Li mass.

The mass measurement is carried out via the determination of the cyclotron
frequency, $\nu_c = {q}{B}/{2\pi}{m}$, 
of ions stored in a homogeneous and stable magnetic field of a Penning trap.
To have access to a wide variety of highly-charged ions an electron beam ion
source (\textsc{Crysis}) in combination with an external ion injector
is used \cite{Beb93}. To produce $^7$Li$^{3+}$ ions, first singly charged Li ions were created in the external ion source 
by evaporating LiBr$_3$ from an oven. The extracted singly charged ions were mass separated and then injected into \textsc{Crysis} for charge breeding. The injection time was $1.43$\,s, the confinement
time, i.e., the time the ions are exposed to the electron impact inside the source, was $20$\,ms and the electron beam energy 14.5\,keV. 
The extracted ion pulse is transported to the double Penning trap system by use of conventional
ion beam optics. Before entering the cylindrical
retardation trap (pre-trap), the ions are charge state selected in a
90$^{\circ}$ double-focusing magnet. 
The pre-trap is used to retard the ions from the transportation energy of
typically $3.4\,q$\,keV to ground potential within $30$\,ms. 
Then the ions are accelerated again to -1\,keV and are transported to the
hyperbolic precision Penning-trap, 
where they are finally retarded to ground potential. An aperture with 1\,mm diameter prevents ions with too large initial radii to enter the precision trap. In this
last stage the trapped ions are subject to an evaporation process by
lowering the trap voltage from $5$ to $0.1$\,V, leaving only the
coldest ions in the trap. In average, not more than $1\!-\!2$ ions
are left in the precision trap after this procedure.

The precision Penning trap is located in the homogeneous magnetic
field of a superconducting solenoid ($B=4.7$\,T). It consists
of a ring electrode and two end-cap electrodes all with hyperbolic
geometry which create an electrostatic quadrupole field.
In these fields the ion's motion can be 
described by three well-defined eigenmotions \cite{Brown86}: an axial motion with frequency $\nu_z$, the so-called magnetron motion with frequency
$\nu_-$, and the modified cyclotron motion with frequency $\nu_+$. The
two radial frequencies obey the relation $\nu_c\ =\ \nu_-\ +\ \nu_+$. 

The ion's cyclotron frequency is probed by exciting the ion's motion by a quadrupolar
radiofrequency signal (rf) and measurement of the time of flight to
the micro-channel-plate detector placed on top of the magnet \cite{konig95,Ber02}. 
Repeating this for different rf frequencies near
the true cyclotron frequency, $\nu_c$, and measuring the time of
flight as a function of the rf frequency, yields a characteristic
time-of-flight cyclotron resonance curve \cite{konig95}. 
In order to obtain the mass from the measured frequency, the
magnetic field has to be calibrated. This is done by the measurement
of the cyclotron frequency, $\nu_c^{\mathrm{ref}}$, of a reference ion
with well-known mass, which is performed almost simultaneously in
order to minimize magnetic-field drifts. 

\begin{figure}
\centering
\resizebox{0.45\textwidth}{!}{%
\includegraphics{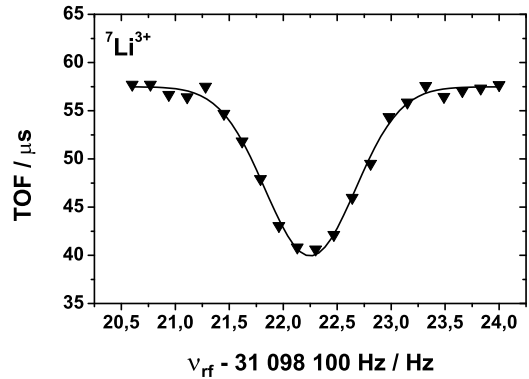}
} 
\caption{The time-of-flight cyclotron frequency resonance of
$^{7}$Li$^{3+}$ from 100 scans representing only $<$10\% of the overall data. 
The central part of the resonance is approximated with a Gaussian (solid
line) in the data evaluation resulting in a FWHM of less than 1\,Hz. The absence of 
well-pronounced sidebands is explained in the text.} 
\label{Fig:timeofflight}
\end{figure}

The mass of the reference ion $m({H_2^+})=2.015\,101\,497\,03(27)$\,u has a relative uncertainty of 0.14\,ppb \cite{Ber02}. $H_2^+$ is produced in the preparation trap by
bombarding the rest gas with 3.4\,keV electrons. The measurements on
$^7$Li$^{3+}$ were performed by using a continuous excitation time
$T_\mathrm{rf}$ of 1\,s. A typical time-of-flight cyclotron
frequency spectrum is shown in Fig.~\ref{Fig:timeofflight}.
The expected sidebands of the resonance \cite{konig95} are supressed. 
This is mainly due to the initial spread in the magnetron radius, since the ions 
are not cooled in the pre-trap, and an incomplete conversion from magnetron to modified cyclotron motion during excitation.

The time-of-flight resonance curve of both, the ion of interest and
reference ion, is measured with 21 equidistant frequency steps
around the center of the resonance frequency. One scan, involving 21 frequency steps,
takes about 40\,s which is repeated twice and after two complete scans the settings were switched between the two ion species; the reference ion $H_2^+$ and the ion of interest $^7$Li$^{3+}$. Switching between ion species takes only
about 1\,s, thus the total cycle time is shorter than 3\,min.
In this way the change in the magnetic field due to temperature or pressure fluctuation between the ion of interest and the reference ion can be reduced.

The mass of the $^7$Li$^{3+}$ is obtained from the observed cyclotron frequency 
ratios of the two ion species:
\begin{equation}
R=\frac{\nu_1}{\nu_2}=\frac{q_1 m_2}{q_2 m_1}\,,
\label{eq:massfromfreq}
\end{equation}
where the subscript 1 denotes the Li ion and subscript 2 the $H_2$ ion.

Since the two frequency measurements are performed in similar ways,
certain systematic uncertainties in the frequency ratio cancel to a
large extent. This is in particular the case for ions which have the
same $q/A$ value \cite{Ber02}. The Li$^{3+}$ ion is close to this
requirement having $q/A=0.43$ compared to $0.5$ for H$_2^+$.
To obtain the atomic mass $m$($^7$Li), one has to correct for the missing $q$
electrons, each with mass $m_e$, and their total binding energy $E_\mathrm{B}$
according to
\begin{equation}
m(^7Li)=m(^7Li^{3+})+qm_e-E_\mathrm{B} \ ,
\label{eq:neutralatommass}
\end{equation}
where $m(^7$Li$^{3+})$ is the experimentally determined ion mass obtained using 
Eq. (\ref{eq:massfromfreq}).
The electron mass is 5.485\,799\,094\,5(24)\,$\times$\,10$^{-4}$\,u with a relative standard uncertainty of 4.4$\times$ 10$^{-10}$ \cite{CODATA02} and since it 
is orders of magnitude smaller than the mass of the $^7$Li, the error introduced by the electron mass can be neglected.

The total electron binding energy is calculated by summing up
the relevant experimental ionization energies tabulated in \cite{Kelly87}. 
For ions with $Z<20$ the relative mass uncertainty from $E_\mathrm{B}$ is $<10^{-11}$.

The results of the three data taking periods are summarized in
Table~\ref{Tab:Freqratio}. The table gives the resulting frequency
ratio $R$ of the ion of interest relative to the reference nuclide as well as the
atomic mass of the studied nuclide, and compares them with the
values given in the literature \cite{AME03}.  
The uncertainty of the $^7$Li mass includes the relative statistical uncertainty (0.4\,ppb) and the relative overall systematic uncertainty (0.52\,ppb). 
The dominant part of the later comes from relativistic effects, ion number dependency, and the $q/A$-asymmetry. The different contributions to the systematic uncertainty are listed in table \ref{Tab:error} and were estimated using the procedures described in ref. \cite{Ber02}.
\begin{table}
\centering \caption{Systematic uncertainty budget for $^7$Li in units of ppb.} 
\label{Tab:error}
\begin{ruledtabular}
\begin{tabular}{llll}
reference mass & 0.18&  binding energy $(E_B)$ & 0.1 \\
relativistic effect & 0.2 & ion nr. dependence & 0.25 \\
$q/A$-asymmetry & 0.33 &contaminating ions & 0.1\\
\hline
&Total systematic & 0.52 &\\
\end{tabular}
\end{ruledtabular}
\end{table}
At the time of the $^6$Li run, we were affected by an uncontrollable internal helium leak in \textsc{Crysis} which had not been present while running $^7$Li.
Since the $^6$Li$^{3+}$ and $^4$He$^{2+}$ are $q/A$ doublets, unwanted $^4$He$^{2+}$ ions were present in the beam, and in the trap mixed together with $^6$Li ions, leading to the large systematic uncertainty in the mass of $^6$Li. 
\begin{table*}
\centering \caption{The measured cyclotron frequency ratio $R$ and the determined  atomic mass $m_{\mathrm{exp}}$, which is compared to the value $m_\mathrm{lit}$ taken from AME2003 \cite{AME03}.  
The error in $R$ is only the statistical error, while $m_{exp}$ includes the systematic uncertainties as well.
} 
\label{Tab:Freqratio}
\begin{ruledtabular}
\begin{tabular}{llll}
&$^{7}$Li& $^{6}$Li & $^{4}$He  \\
\hline
$R$ & 0.861\,847\,167\,21(31)& 1.005\,292\,631\,83(80) & 1.007\,171\,503\,45(53)\\
$m_\mathrm{exp}$ & 7.016\,003\,425\,6\,(45)\,u & 6.015\,122\,890\,(40)\,u &4.002\,603\,253\,3(26)\,u\\   
$m_\mathrm{lit}$ &	7.016\,004\,550\,(80)\,u &6.015\,122\,795\,(16)\,u&4.002\,603\,254\,15(16)\,u\\
\end{tabular}
\end{ruledtabular}
\end{table*}

\begin{figure}
\centering
\resizebox{0.50\textwidth}{!}{%
\includegraphics{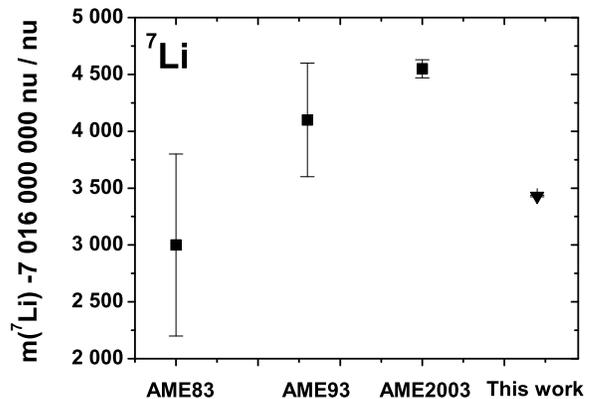}
}
\caption{Comparison of the atomic mass value of $^7$Li from the measurement
reported here with previous results published in the atomic-mass
evaluations AME83 \cite{AME:85}, AME93 \cite{AME:93}, and AME2003 \cite{AME03}. The recent AME2003 value deviates by 160\,ppb from our measurement. }
\label{Fig:result}       
\end{figure}

A comparison of our result for the $^7$Li atomic mass with previous
data is shown in Fig.~\ref{Fig:result}. The AME83 \cite{AME:85} value of
$^7$Li is based on a reaction energy and has an uncertainty of 0.8\,$\mu$u. Similarly, the
AME93 \cite{AME:93} value, which is derived from $^7$Li(p,n)$^7$Be reaction $Q$-value, has an uncertainty of 0.5\,$\mu$u. 
The most recent AME2003 value has a much reduced uncertainty of only 0.08\,$\mu$u. In this case the mass of $^7$Li has been derived using as input data the mass of $^6$Li and the $^6$Li(n,$\gamma$)$^7$Li reaction $Q$-value \cite{AME03}. 
The $^7$Li mass value from AME2003 deviates significantly $(>1\,\mu$u) from our result,
which means that at least one of the two input data used to derive the $^7$Li mass must be wrong.
 
\begin{figure}
\centering
\resizebox{0.50\textwidth}{!}{%
\includegraphics{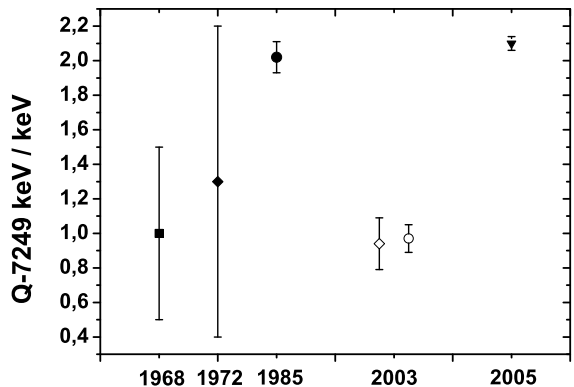}
}
\caption{The $Q$-value of the $^6$Li(n,$\gamma$)$^7$Li reaction in chronological order.
The values are 1968 from ref. \cite{Spilling:68}, 1972 from ref. \cite{Kamp:72}, 1985 from ref. \cite{kok:85}. The first point at 2003 is from ref. \cite{Firestone:03} and the second from ref. \cite{AME03}. The 2005 point is the value derived from our mass measurements using $^6$Li and $^7$Li.}
\label{Fig:Q-value}       
\end{figure}

Different $Q$-values exist in the literature \cite{Spilling:68,Kamp:72,kok:85,Firestone:03}, see fig. \ref{Fig:Q-value}.
Note, that the $Q$-value from 1985 \cite{kok:85} deviates by about 1\,keV from the value in AME2003 \cite{AME03} which is claimed to be based upon recalibrated data from ref. \cite{kok:85} and the recent data from ref. \cite{Firestone:03}. Furthermore, the work in ref. \cite{Firestone:03} is not published.  
The mass of $^6$Li is known to 2.7\,ppb uncertainty \cite{Dunn01}; however, to shed light upon this large deviation we have measured the mass of the $^6$Li and found an agreement within 2.4 $\sigma$ compared to the literature value (Table \ref{Tab:Freqratio}).
Using our mass values for $^6$Li and $^7$Li reported here, a $Q$-value of 7251.10(4)\,keV is derived. For the 2003 $^7$Li mass calculation a $Q$-value of 7249.97(8)\,keV \cite{AME03} has been used which deviates by more than 1\,keV from our value and can explain the large discrepancy observed. Note, that the $Q$-value derived from our mass measurement is in agreement with the $Q$-value from Ref. \cite{kok:85} of 7251.02(9)\,keV.  
 
The excellent agreement of our simultaneously measured $^4$He mass with the literature value gives further confidence in the $^7$Li mass value reported here, where both measurements are at exactly the same level of precision. 

Summarizing, the result of a high-accuracy atomic mass
measurement of $^7$Li with the Penning trap mass spectrometer
\textsc{Smiletrap} has been reported. The mass of $^7$Li was measured
directly with unprecedented accuracy and the result has been compared to
previous published mass values. The new atomic-mass that was determined has a deviation of 1.1\,$\mu$u as compared to the AME2003 value which seems to be due to a wrong $^6$Li(n,$\gamma$)$^7$Li reaction $Q$-value used in the latest atomic mass evaluation \cite{AME03}. The new high-accuracy mass value for $^7$Li is an
important input parameter for transition isotope shift and nuclear
charge radii measurements of the Li isotopes  \cite{noer05,Bush03,Ewald04}. 
It can also be used as reference mass for calibration purposes in high-accuracy Penning trap mass spectrometry of short-lived nuclides. Furthermore, in the evaluation of the masses of $^7$Be and $^8$Li, 
the mass of $^7$Li is used as input parameter \cite{AME03}.

\begin{acknowledgments} 
We gratefully acknowledge support from the Knut and
Alice Wallenberg Foundation, the European R\&D network HITRAP (contract No. HPRI CT 2001 50036), and from the Swedish research council VR. 
One of the authors (K. B.) acknowledges support by the Helmholtz Association
of National Research Centres (HGF) under contract No. VH-NG-037. 
We are also indebted to the Manne Siegbahn Laboratory for the support.

\end{acknowledgments}
%


\end{document}